\documentclass[onecolumn,superscriptaddress,aps,floatfix,12pt]{revtex4-2}
\usepackage{hyperref}
\usepackage{amsmath}
\usepackage{amssymb}
\usepackage{graphicx}
\usepackage{mathrsfs}
\usepackage{braket}
\usepackage{bm,bbm}
\usepackage{color}

\usepackage{amsmath}
\usepackage{graphicx}

\newcommand{\x}{\langle{x}\rangle}
\newcommand{\xx}{\langle{x^2}\rangle}
\newcommand{\br}{\bm{r}}
\newcommand{\bk}{\bm{k}}
 
\begin{document}
\title{Anomalous Quantum Propagation of Microcavity Exciton Polaritons}

\author{Tian Lingyu}
\affiliation{Beijing Academy of Quantum Information Sciences, Beijing, 100193, P.R. China}

\author{Yutian Peng}
\affiliation{State Key Laboratory of Low-Dimensional Quantum Physics and Department of Physics, Tsinghua University, Beijing, 100084, P.R. China}
\affiliation{Beijing Academy of Quantum Information Sciences, Beijing, 100193, P.R. China}

\author{Qihua Xiong}
\email{qihua{\_}xiong@tsinghua.edu.cn}
\affiliation{State Key Laboratory of Low-Dimensional Quantum Physics and Department of Physics, Tsinghua University, Beijing, 100084, P.R. China}
\affiliation{Beijing Academy of Quantum Information Sciences, Beijing, 100193, P.R. China}
\affiliation{Frontier Science Center for Quantum Information, Beijing, 100084, P.R. China}
\affiliation{Collaborative Innovation Center of Quantum Matter, Beijing 100084, P.R. China}

\author{Sanjib Ghosh}
\email{sanjibghosh@baqis.ac.cn}
\affiliation{Beijing Academy of Quantum Information Sciences, Beijing, 100193, P.R. China}

% %
\date{\today}
% %

%%%%%%
\begin{abstract}
Here, we explore the quantum propagation of exciton polaritons in semiconductor microcavities, exhibiting intriguing effects such as interactions, decay, and disorder scatterings. Our investigation uncovers anomalies in their quantum propagation, deviating from predictions based on existing theories. By applying scaling theory, we elucidate the true nature of exciton polariton propagation, unveiling a localization phase that characteristically differs from Anderson localization. Our numerical results agree with the self-consistent theory developed for exciton polariton condensates, incorporating non-linearity and finite lifetime.
\end{abstract}
\maketitle

%%%%%%%%%%%%%%%%%%%%%%
\section{Introduction} 
Quantum propagation is a complex phenomenon arising from the intricate interplay of coherent interference effects, impurity scatterings, and inter-particle interactions. Traditionally, theories of quantum propagation have been developed in the context of electronic systems, where various transport properties are analyzed~\cite{Akkermans_Montambaux_2007}. However, with the advent of the second quantum revolution~\cite{deutsch2020harnessing}, access to new types of quantum systems, such as Bose-Einstein condensates~\cite{anglin2002bose,chen2022continuous,kasprzak2006bose}, superconducting qubits~\cite{houck2012chip}, and trapped ions~\cite{blatt2012quantum}, has become available. In these systems, probing intriguing phenomena like quantum propagation is no longer limited to measuring indirect properties like conductivity or resistivity; instead, there is direct access to the distribution of particles in space and time~\cite{jendrzejewski2012three,balasubrahmaniyam2023enhanced,chen2023unraveling}. These new-generation quantum systems offer the potential to realize a richer form of quantum propagation beyond the confines of traditional transport theories.

Exciton polaritons in microcavities, as bosonic quasi-particles blending properties of light and matter~\cite{carusotto2013quantum,byrnes2014exciton}, exhibit ultrafast dynamics~\cite{mrejen2019transient,xu2023ultrafast,ramezani2019ultrafast}, small effective masses~\cite{pinsker2017effects}, and enhanced nonlinearity~\cite{EstrechoPRB2019,wu2021perovskite}. Furthermore, their short lifetimes and external excitation can drive the system into a highly non-Hermitian state, allowing particles to enter or leave depending on the balance between loss and gain~\cite{Hanai2019PRL}. Possessing all these attractive features, exciton polaritons can undergo Bose-Einstein condensation above a threshold~\cite{kasprzak2006bose}, making them versatile platforms for exploring various quantum effects~\cite{ghoshPI22}. Certainly, over the decades, the field of exciton polaritons has matured, demonstrating remarkable advancements in their creation, manipulation, and detection. For example, polariton condensates have been realized, propagated, controlled, and imaged in various systems~\cite{wertz2012propagation,hu2017imaging,mrejen2019transient}, including those at elevated temperatures~\cite{Su2017,Su2020,su2018room,chen2023unraveling}. From a practical standpoint, these capabilities are crucial for information processing~\cite{opala2019neuromorphic,banerjee2020coupling,nigro2022integrated,xu2021superpolynomial}, requiring data transport via propagating polariton condensates in semiconductor microcavities.
%From a practical standpoint, these capabilities are crucial for information processing schemes~\cite{opala2019neuromorphic,banerjee2020coupling,nigro2022integrated,xu2021superpolynomial}, requiring the transportation of data via propagating polariton condensates from one point of a microcavity to another.

Inevitably, one would like to know whether polaritons can actually traverse to carry information between two distant points within an elongated microcavity. For an accurate understanding, it is crucial to consider realistic effects such as finite lifetimes, disorder scattering, and nonlinear interactions. While a finite lifetime itself introduces decay that can hinder long-range propagation, it can be compensated by increasing the polariton population. More fundamentally, one needs to find if polaritons can travel long distances before they all leave the system. Unlike classical particles, exciton polaritons can exhibit an intricate form of propagation due to the quantum superposition of simultaneous multiple scattering events, resulting in complex interference effects~\cite{lee1985disordered}. Moreover, nonlinear interactions between exciton polaritons can contribute additional complexities to their quantum propagation, paving the way for even more subtler effects to emerge. The existing concepts developed for low-temperature electronic systems (described by the Schr\"odinger's equation) or cold atomic systems (described by the Gross-Pitaevskii equation) are not directly applicable to exciton polaritons. A microscopic theory for the quantum propagation of exciton polaritons needs to be developed.

Aligned with the physical scenario, here we delve into quantum propagation in a one-dimensional (1D) geometry, representing an elongated microcavity for condensate transport. Employing a scaling theory, we developed a systematic approach to analyze the dynamics of polariton wave packets and uncover their fate at long times. Explicit comparisons are drawn between these findings and the behaviors exhibited by particles governed by Schr\"odinger's and Gross-Pitaevskii equations, highlighting certain anomalous characteristics.

%%%%%%

\section{Basic theory} 
The propagation of quantum particles is explored across diverse systems, spanning solid-state systems, ultracold atoms, and optical setups. For our study, we categorize these particles broadly as Schr\"odinger and Gross-Pitaevskii particles, based on their respective equations of motion. To encompass all these particle types, we consider a generalized quantum equation~\cite{carusotto2013quantum,wouters2007excitations}:
\begin{eqnarray}
i\hbar \dot{\psi}(\bm{r},t) = \left[ (H-i\gamma/2) + \alpha |\psi(\bm{r},t)|^2 \right]\psi(\bm{r},t)
\label{GLE}
\end{eqnarray}
where $\psi(\br,t)$ is the wave function at point $\br$ and time $t$, $H$ is the Hamiltonian, $\gamma$ represents the decay and $\alpha$ is the nonlinearity strength. Depending on the presence of decay $\gamma$ and nonlinearity $\alpha$, Eq.~\ref{GLE} represents Schr\"{o}dinger particles ($\gamma =0=\alpha$), Gross-Pitaevskii (GP) particles ($\gamma =0$ and $\alpha \ne 0$), and  exciton polariton condensates ($\gamma \ne 0$ and $\alpha \ne 0$). 

The Hamiltonian $\hat{H} = T(\hat{\bk}) + V(\bm{r}) $ includes a kinetic part $T(\hat{\bk})$ and a potential part $V(\bm{r})=V_0 R(\bm{r})$ which represents disorder in the system. $R(\bm{r})$ is a Gaussian random variable and $V_0$ is the strength of disorder ($V_0 = 3 \,\, meV$). The kinetic energy $T(\bm{k})= [E_x+E_c(\bm{k})-\sqrt{[E_x-E_c(\bm{k})]^2 + \Omega^2}\, ]/2$ which represents the lower polariton branch with the exciton energy $E_x$, the cavity photon dispersion $E_c(\bm{k})$, and the vacuum Rabi splitting $\Omega$ (see supplementary information for details). For $\gamma >0$, the total intensity $A_0(t) = \int d\br |\psi (\bm{r},t)|^2$ show trivial exponential decay $\propto \exp({-\gamma t/\hbar})$. However, such an overall decay does not reveal any information about the spatial distribution of intensity. Focusing on the spatial propagation, we normalise the intensities $|\psi (\bm{r},t)|^2/A_0(t) $ to maintain a total of $1$. Finally, the average intensity distribution $I(\br,t)$ is given by: 
\begin{eqnarray}
I(\br,t) = \frac{1}{N} \sum_j \left[ {|\psi (\bm{r},t)|^2}/{A_0(t)} \right]_j
\end{eqnarray}
where $N$ is the number of disorder realizations ($N=25000$ for all our results) and $j$ indicates an individual disorder realization. The average expectation value of an observable $O$ is defined as $\langle O \rangle = \int d\br O(\br) I(\br,t)$. Furthermore, since the nonlinearity in Eq.~\ref{GLE} relies on the term $\alpha|\psi(\br,t)|^2$, the effective nonlinear interaction can be altered by adjusting either $\alpha$ or the intensity. To circumvent this redundancy, we define the initial interaction energy as $E_\alpha = \alpha A_0(t=0)$ instead of directly using $\alpha$.

\begin{figure}[h]
\includegraphics[width=1\linewidth]{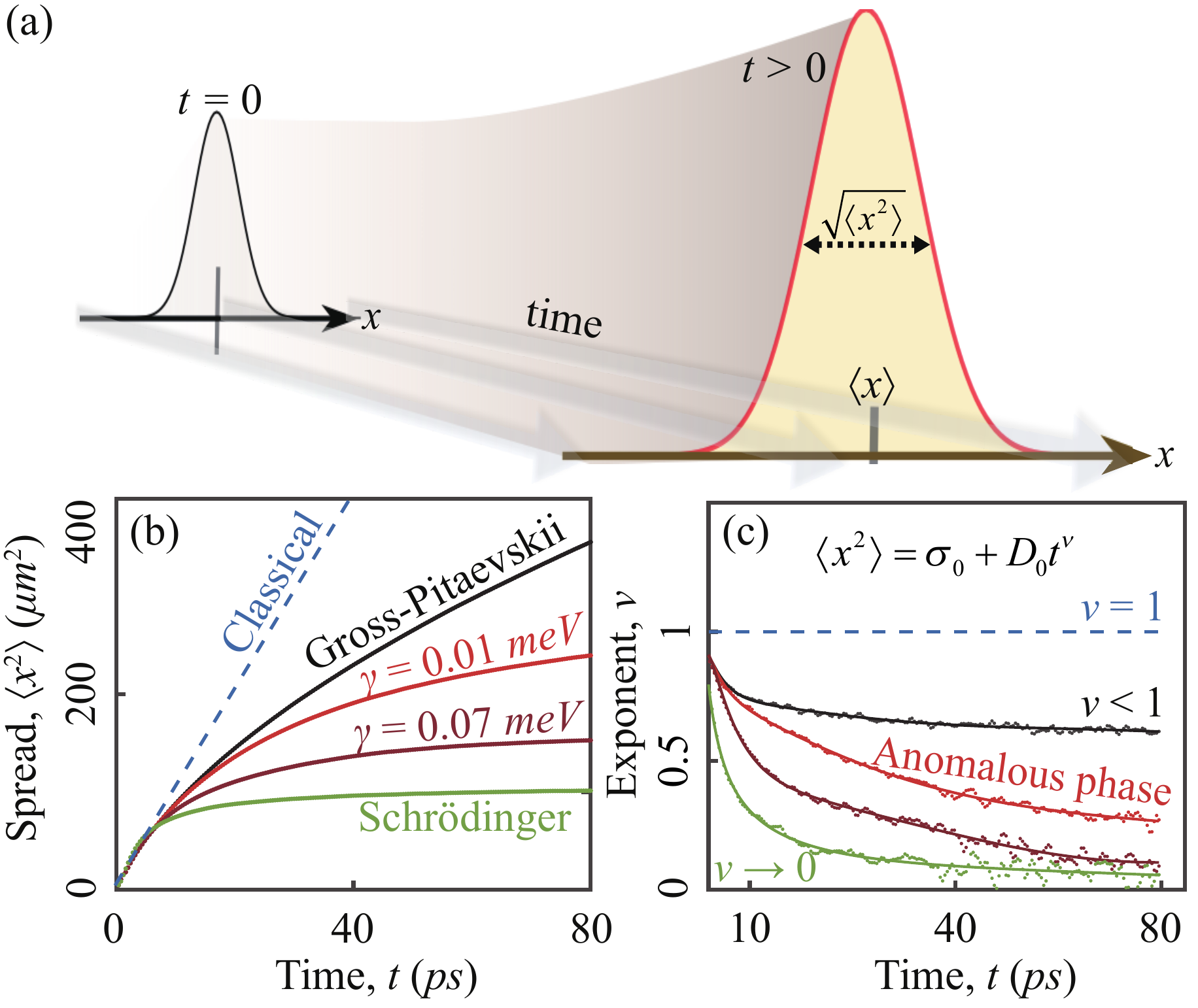}
\caption{Quantum propagation dynamics of microcavity exciton polaritons. (a) Spreading of a wave packet which was launched at $t=0$ with a spread $\xx=\sigma_0$ and center of mass $\x = 0$ in a disordered system. (b) and (c) represent the spread $\xx$ and the diffusion exponent $\nu$ as functions of propagation time $t$ for different types of particles, Schr\"odinger  (green), GP (black), classical (dotted blue) and  exciton polaritons with different decays [red ($\gamma =0.01\,meV$) and dark-red ($\gamma =0.07\,meV$)] respectively. Here, $E_{\alpha} = 26.6 \, meV $ for GP particles and exciton polaritons.}
\label{Spreading}
\end{figure}

\section{Wave-packet spreading} 
In our consideration, at time $t=0$ a wave packet $\psi(\br,0)=\exp[-x^2/(2\sigma_0) + ik_0 x]$ is launched with a spatial spread ${\langle x^2 \rangle} =\sigma_0=9 \, \mu m^2$ and a center of mass $\langle x\rangle =0$. Such an initial wave packet can be prepared with a resonant coherent laser pulse. As schematically represented in Fig.~\ref{Spreading}a,  the wave packet evolves over time in a disordered system to exhibit a new spatial spread $\xx$ and center of mass $\x$. Fig.~\ref{Spreading}b shows $\xx$ as a function of time for Schr\"odinger and GP particles and exciton polaritons. 

In disordered systems, long-time propagation is primarily influenced by multiple scattering events. In classical scenarios, these events incoherently cause the direction of propagation to be randomly distributed, resulting in overall diffusive motion. However, for quantum particles, these scattering events induce complex interference effects that counteract incoherent diffusion~\cite{ghosh2014coherent}. When interference effects are strong enough, diffusion can be entirely suppressed, leading to Anderson localization of quantum particles~\cite{anderson1958absence}. As a consequence of this localization process, the spreading of Schr\"odinger particles (see Fig.~\ref{Spreading}b) rapidly decelerates and halts at a finite value $\sim\xi^2$, where $\xi$ represents the localization length. The localization of particles occurs when eigenstates are confined within specific regions of a system. While this description aptly applies to Schrödinger particles, GP particles experience nonlinear interactions, enabling the coupling of different localized eigenstates. This coupling allows GP particles to hop from one localized state to another, resulting in a propagation akin to diffusion, albeit at a slower rate (sub-diffusion). As depicted in Fig.~\ref{Spreading}b, GP particles continue to spread over time, indicating sub-diffusion due to the destruction of Anderson localization by nonlinear interactions~\cite{Shepelyansky08}.

Considering the intricate interplay between quantum coherence and nonlinear interactions observed in GP particles, it is intriguing to explore whether the finite lifetime of microcavity exciton polaritons can induce novel effects in quantum propagation. Fig.~\ref{Spreading}b shows that exciton polaritons spread faster than Schr\"odinger particles but slower than GP particles. However, it is unclear if such motion is merely another form of sub-diffusion.

\section{Diffusion exponent} 
For a more quantitative analysis, we calculate the exponent $\nu$ of the diffusion relation $\langle x^2 \rangle = \sigma_0 + D_0 t^\nu$. For classical diffusion the exponent $\nu=1$ signifying a constant linear spread of particles with propagation time $t$. Schrodinger particles undergo Anderson localization with an exponent $\nu\to 0$ indicating the halt of diffusion. For GP particles, the calculated exponent $0<\nu<1$ signifies a slow but steady diffusion (sub-diffusion). 

Intriguingly, exciton polaritons exhibit a prolonged time-dependent diffusion exponent $\nu$ until a timescale where both Schr\"odinger and GP particles stabilize to their asymptotic phase. However, eventually the exponent approach towards the localization limit $\nu\to 0$, even though  we have considered the same non-linear parameter both for GP particles and exciton polaritons.

\begin{figure}[h]
\includegraphics[width=1\linewidth]{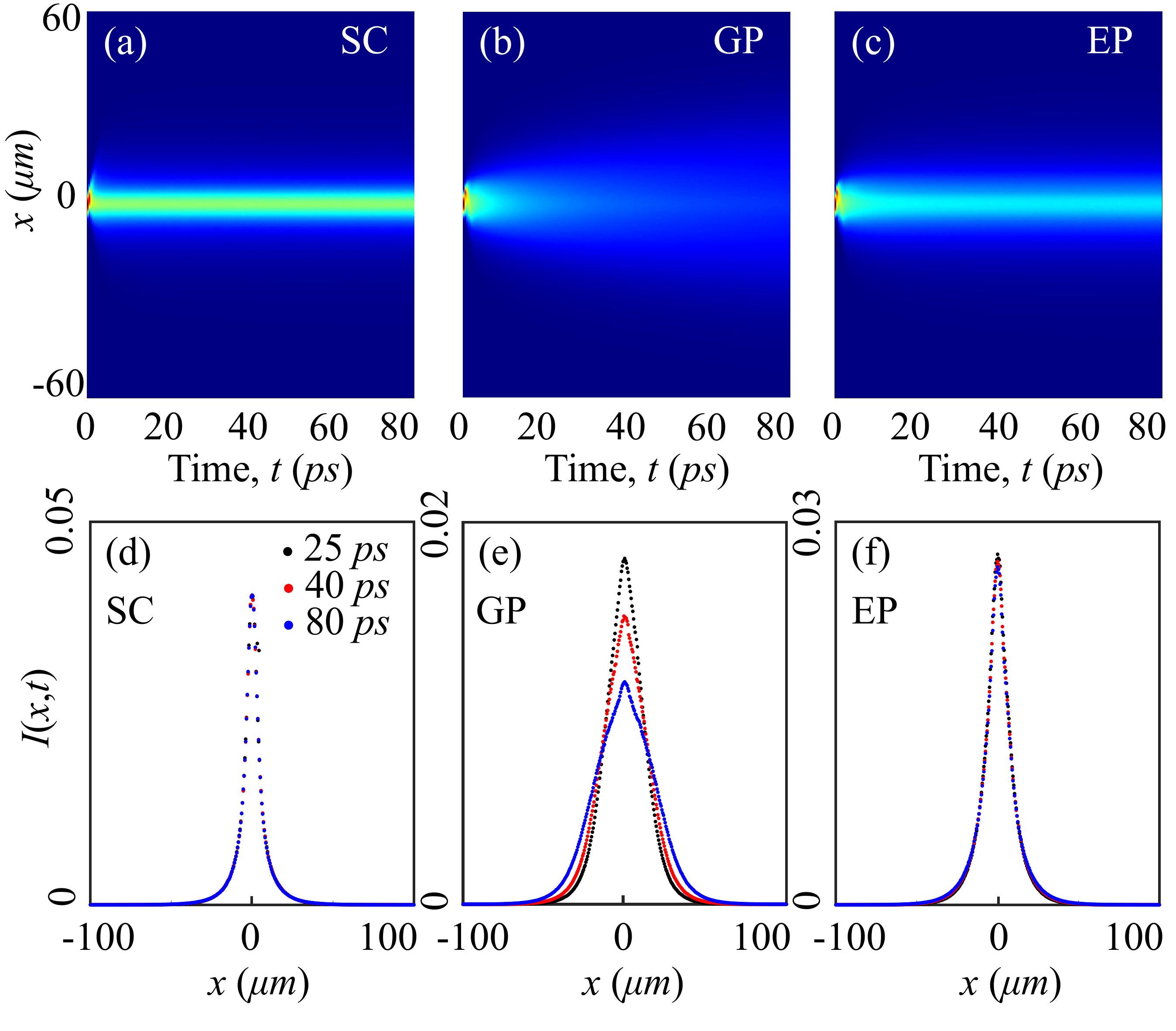}
\caption{Intensity distribution. (a-c) Averaged intensity distribution $I(x,t)$ for Schr\"odinger (SC) and GP particles and exciton polaritons (EP) as a function of time and space, which visually show how wave packet propagates. (d-f) We show $I(x,t)$ at three different times $t = 25 \, ps$ (black), $40 \, ps$ (red) and $80 \, ps$ (blue) for three types of particles. Clearly, while GP particles are expanding with time, exciton polaritons and Schr\"odinger particles are localised with different localization lengths. Here, $E_{\alpha} = 26.6 \, meV $ for GP particles, and $E_{\alpha} = 26.6 \, meV $ and $\gamma = 0.07 \, meV$ for exciton polaritons.}
\label{Idistribution}
\end{figure}

\section{Intensity distribution} 
For direct comparison, we analyze the intensity distributions of Schr\"odinger and GP particles alongside exciton polaritons in Fig.~\ref{Idistribution}. We observe that the intensity distribution, $I(\br,t)$, for Schr\"odinger particles ceases to spread (due to Anderson localization) after an initial evolution (see Fig.~\ref{Idistribution}a \& d). In contrast, $I(\br,t)$ for GP particles, depicted in Fig.~\ref{Idistribution}b \& e, continually expands (due to sub-diffusion) over time. Fig.~\ref{Idistribution}c \& f reveal that $I(\br,t)$ for exciton polaritons exhibits a localization behavior similar to Schr\"odinger particles. However, it is evident that the spatial extend of exciton polaritons exceeds that of Schr\"odinger particles. A closer look is required to understand the true nature of exciton polariton propagation.

\section{Scaling theory} 
We utilize scaling theory of localization~\cite{GangOf4}, for a comprehensive analysis. This theory explores the Thouless parameter $G$, defined as the ratio $G = \Delta E/\delta E$, where $\Delta E = \hbar D/L^2$ is the Thouless energy, $\delta E= 1/[\rho L^d]$ is the energy level spacing, $D$ is the diffusion coefficient in a system of size $L$, $\rho$ is the density of states, and $d$ is the dimension~\cite{thouless1974electrons}. Scaling theory examines the evolution of $G$ as the system transitions from microscopic to macroscopic scales.

\begin{figure}
\includegraphics[width=1\linewidth]{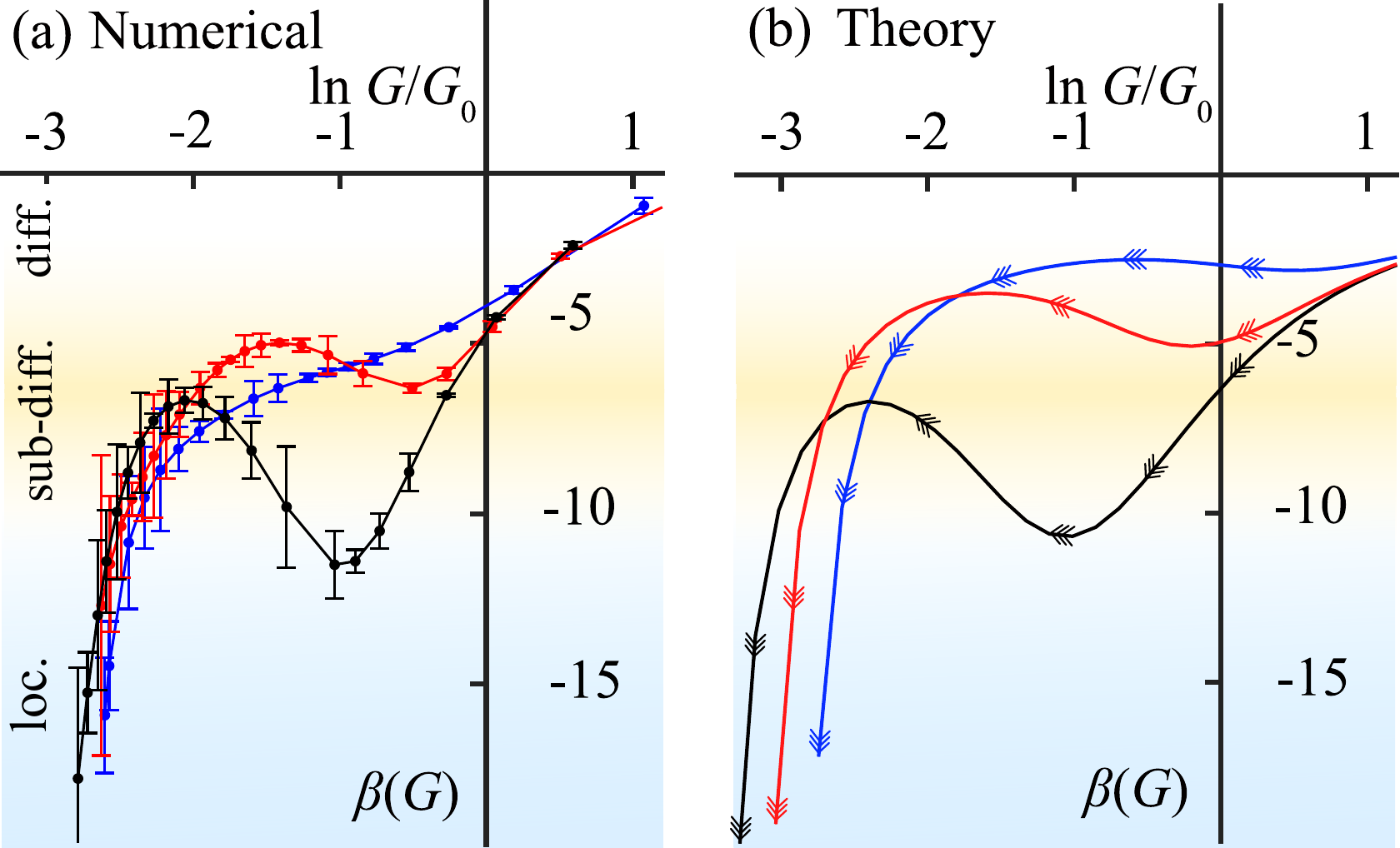}
\caption{Scaling function for the microcavity exciton polariton propagation. (a) Numerically obtained scaling function $\beta$ as a function of Thouless parameter $G$. (b) Scaling function obtained from the self-consistent theory. The arrows indicate the flow of scaling function during the propagation. Three curves correspond to $E_{\alpha} = 8 \, meV$ (blue), $2.7 \, meV$ (red) and $0.8 \, meV$ (black) with a decay $\gamma = 0.01 \, meV$.}
\label{Beta}
\end{figure} 

In a dynamic scenario, the effective system size is determined by the width of the wave packet $ L_t = \sqrt{\xx}$, which evolves from an initial small value to a larger one over time. The Thouless energy at time $t$ is given by $\Delta E = \hbar/t$, as $L_t$ relates to $D$ as $L_t^2 \sim Dt $. The energy level spacing corresponding to $L_t$ is $\delta E=1/[\rho (L_t)^d]$. These relations lead to the Thouless parameter $ G = \hbar \rho t^{-1} \xx^{d/2}$~\cite{cherroret2014nonlinear}. The corresponding scaling function is given by,
\begin{eqnarray}
    \beta(G)= \frac{d \ln G(L_t)}{d\ln L_t}.
\end{eqnarray}
Fig.~\ref{Beta}a and b shows the numerically and theoretically calculated scaling functions for exciton polaritons. The nature of propagation is identified from the flow of scaling function during the course of time. While a flow towards a finite value indicates a propagating phase (such as diffusion or sub-diffusion), a localized phase is identified by a flow towards $\beta \to -\infty$~\cite{GangOf4}. 

In a purely diffusive system, $\xx \simeq D_0 t$ which implies $L_t\simeq \sqrt{D_0t}$, $G\sim  L_t^{d-2}$ and $\beta \simeq d-2$. When a wave packet is launched with a narrow width, initially ($l_0\ll L_t < \xi$), it diffusively spreads (see Fig.~\ref{Spreading}b), regardless of the type of particles involved. Thus, one expects $\beta$ to start from the value $\sim d-2$ at small times. Indeed, our calculated scaling function starts near $\beta \simeq -1$ (for our 1D system) and then flows towards negative values. However, it quickly transitions to a sub-diffusion phase (indicated by regions shaded in yellow in Fig.~\ref{Beta}). We observe that this re-entrance to sub-diffusion is prominent for weaker nonlinearity. It is crucial to note that without observing the propagation dynamics for a sufficiently long time, one might erroneously conclude that the fate of exciton polariton propagation resembles that of GP particles, exhibiting sub-diffusion. At longer propagation times, we observe that the scaling function eventually flows back towards large negative values (regions with blue shading in Fig.~\ref{Beta}). Such a flow indicates an asymptotic localization phase. In this phase, the spreading of the wave packet is limited by a localization length $\xi_\text{ep}$, implying $\xx \to {\xi_\text{ep}}^2$ and $\beta \to -\infty$.

Next, we apply the self-consistent theory to theoretically evaluate the scaling function~\cite{vollhardt1982scaling}. In this theory, quantum propagation seen as a combination of incoherent diffusion and coherent interference. The diffusion is described by a process where $\psi$ and $\psi^*$ follow the same scattering path. The interference term, known as the Cooperon, is described by a configuration where $\psi$ and $\psi^*$ follows the same path in opposite directions. A direct application of this theory (applicable to Schr\"odinger particles) predicts a monotonic scaling function that disagrees with Fig.~\ref{Beta}a. A modified version of this theory was developed for GP particles by including an effective dephasing term ($\sim\alpha |\psi|^2$) in the Cooperon~\cite{cherroret16}. This theory explains the sub-diffusion of GP particles. Here, we adapted a similar strategy to include the effect of decay and non-linearity simultaneously (see Supplementary Information). Our theoretical result (Fig.~\ref{Beta}b) qualitatively agrees with the localization phase seen in our numerical results (Fig.~\ref{Beta}a). However, it is unsettled if this localization phenomenon is identical to the Anderson localization. 

\begin{figure}
\includegraphics[width=1\linewidth]{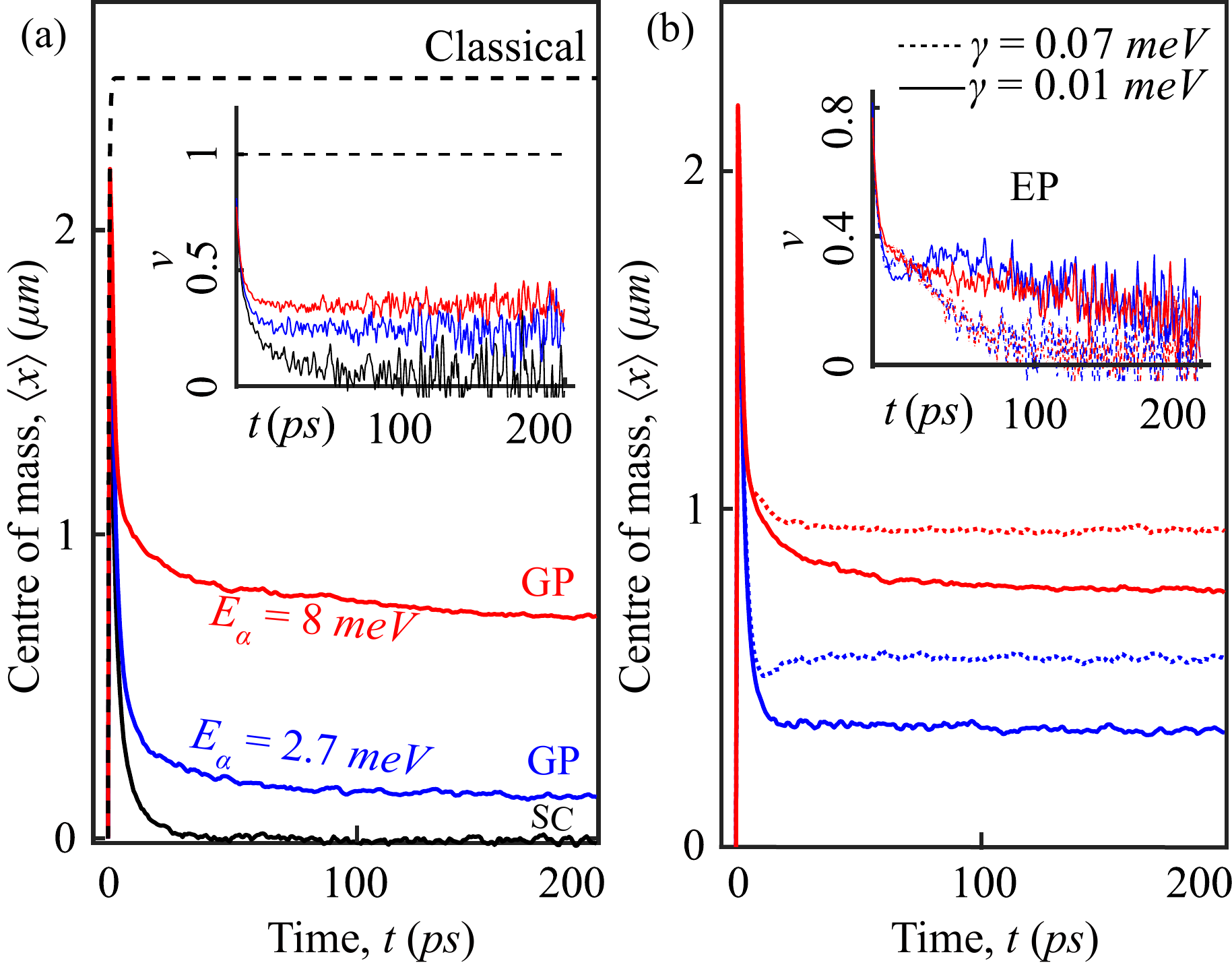}
\caption{Center of mass $\langle x \rangle$ dynamics. (a) $\x$ for Schrodinger (black) and GP particles (blue and red for $E_\alpha=2.7\, meV$ and $8\, meV$) as a function of time. While for Schrodinger particles $\x \to 0$ as time increases, for GP particles $\x$ remains nonzero at all times. The diffusion exponents $\nu$ (inset) shows different behaviour for Schr\"odinger ($\nu\to 0$) and GP particles ($\nu\ne 0$). (b) Anomalous center of mass motion for exciton polaritons with $\gamma = 0.07\,meV$ (dashed curves) and $0.01 \, meV$ (solid curves), which remain nonzero even though $\nu \to 0$ (inset). We consider $E_{\alpha}=2.7 \, meV$ (blue solid and dotted curves) and $8\,meV$ (red solid and dotted curves) for both (a) and (b).}
\label{CM_motion}
\end{figure}

\section{Center of mass motion} 
We observed explicit anomalous features in the center of mass motion of exciton polaritons. Generally, the dynamics of the center of mass, $\x$, reflects the microscopic nature of quantum propagation. Classically, particles with an initial momentum $\bk_0$ exhibit ballistic motion, wherein their average position $\x$ increases linearly with time, up to the scattering mean free path  $l_0$. Beyond $l_0$, the propagation direction becomes randomized due to disorder scatterings resulting in a zero average velocity, implying no further increase in $\x$. This is described by the classical Drude model, where the average velocity $v$ satisfies $\dot{v} = -v/\tau_0$ ($\tau_0$ is the mean free time) with the solution $\x = l_0 (1-e^{-t/\tau_0})$. From the results shown in Fig.~\ref{CM_motion}a, we find $l_0\approx 2.5\, \mu m$ and $\tau_0\approx 0.35\, ps$. 

The classical picture fails to describe Schr\"{o}dinger particles, where quantum interference effects become significant. Here, after the initial ballistic propagation, the center of mass $\x$ starts moving backwards to ultimately reach the initial position $\x =0$ (see Fig.~\ref{CM_motion}a) with the diffusion exponent $\nu\to 0$ (see inset to Fig.~\ref{CM_motion}a). This behaviour of Schr\"{o}dinger particles is directly related to Anderson localization~\cite{prat2019quantum,sajjad2022observation}. GP particles undergo sub-diffusion through the destruction of Anderson localization due to nonlinear interaction. The absence of Anderson localization in GP particles~\cite{janarek2020quantum} manifests as a nonzero center of mass $\x > 0$ and diffusion exponent $\nu>0$ for all times as can be seen in Fig.~\ref{CM_motion}a . 

In our numerical result shown in Fig.~\ref{CM_motion}b, we find that $\x$ for exciton polaritons remains nonzero for all times. Such a finding would indicate a sub-diffusive propagation of exciton polaritons like GP particles. Indeed, exciton polaritons interact among themselves through the nonlinear interaction and thus it would be natural to anticipate the destruction of Anderson localization and the restoration of a diffusion-like propagation. However, while closely examining the diffusion exponent $\nu$ from the spatial spread $\xx$ in Fig.~\ref{CM_motion}b, we find exactly the opposite behaviour, as already indicated in Fig.~\ref{Spreading}c. We find that $\nu$ steadily approach the localization limit $\nu \to 0$ (inset to Fig.~\ref{CM_motion}b) confirming the conclusion obtained from the scaling function in Fig.~\ref{Beta}. This shows an elusive nature of the quantum exciton polariton propagation, where the localised phase confirmed by the scaling theory shows a different physical character than Anderson localization of Schr\"odinger particles.

%\section{Discussion}

\section{Summary} 
In summary, our study reveals anomalous properties in the quantum propagation of microcavity exciton polaritons contrasting with existing theories for Schr\"odinger and Gross-Pitaevskii particles. We have identified a distinct localization phase governing the long-time propagation of exciton polaritons, which differs from Anderson localization observed in Schr\"odinger particles. Through detailed numerical analysis and the calculation of the scaling function, we have characterized this localization phase and developed a self-consistent theory that agrees well with our numerical findings. Moreover, our investigation provides clear evidence, particularly from the center of mass motion, supporting the notion that this localization phenomenon of exciton polaritons exhibits unique characteristics compared to Anderson localization. This study underscores the complexity and richness of quantum exciton polariton propagation, offering insights into a novel physical phenomenon.

\section{Acknowledgments}
S. G. gratefully acknowledges funding support from the Excellent Young Scientists Fund Program (Overseas) of China, and the National Natural Science Foundation of China (Grant No. 12274034). Q. X. gratefully acknowledges funding support from the National Natural Science Foundation of China (Grant No. 12020101003, 92250301, and 12250710126) and strong support from the State Key Laboratory of Low-Dimensional Quantum Physics at Tsinghua University.
%%%%%%%%%%%%%%%%%%%%%%%%%%%%%
\bibliographystyle{apsrev4-2}
\bibliography{referencev2.bib}

\begin{thebibliography}{42}%
\makeatletter
\providecommand \@ifxundefined [1]{%
 \@ifx{#1\undefined}
}%
\providecommand \@ifnum [1]{%
 \ifnum #1\expandafter \@firstoftwo
 \else \expandafter \@secondoftwo
 \fi
}%
\providecommand \@ifx [1]{%
 \ifx #1\expandafter \@firstoftwo
 \else \expandafter \@secondoftwo
 \fi
}%
\providecommand \natexlab [1]{#1}%
\providecommand \enquote  [1]{``#1''}%
\providecommand \bibnamefont  [1]{#1}%
\providecommand \bibfnamefont [1]{#1}%
\providecommand \citenamefont [1]{#1}%
\providecommand \href@noop [0]{\@secondoftwo}%
\providecommand \href [0]{\begingroup \@sanitize@url \@href}%
\providecommand \@href[1]{\@@startlink{#1}\@@href}%
\providecommand \@@href[1]{\endgroup#1\@@endlink}%
\providecommand \@sanitize@url [0]{\catcode `\\12\catcode `\$12\catcode
  `\&12\catcode `\#12\catcode `\^12\catcode `\_12\catcode `\%12\relax}%
\providecommand \@@startlink[1]{}%
\providecommand \@@endlink[0]{}%
\providecommand \url  [0]{\begingroup\@sanitize@url \@url }%
\providecommand \@url [1]{\endgroup\@href {#1}{\urlprefix }}%
\providecommand \urlprefix  [0]{URL }%
\providecommand \Eprint [0]{\href }%
\providecommand \doibase [0]{https://doi.org/}%
\providecommand \selectlanguage [0]{\@gobble}%
\providecommand \bibinfo  [0]{\@secondoftwo}%
\providecommand \bibfield  [0]{\@secondoftwo}%
\providecommand \translation [1]{[#1]}%
\providecommand \BibitemOpen [0]{}%
\providecommand \bibitemStop [0]{}%
\providecommand \bibitemNoStop [0]{.\EOS\space}%
\providecommand \EOS [0]{\spacefactor3000\relax}%
\providecommand \BibitemShut  [1]{\csname bibitem#1\endcsname}%
\let\auto@bib@innerbib\@empty
%</preamble>
\bibitem [{\citenamefont {Akkermans}\ and\ \citenamefont
  {Montambaux}(2007)}]{Akkermans_Montambaux_2007}%
  \BibitemOpen
  \bibfield  {author} {\bibinfo {author} {\bibfnamefont {E.}~\bibnamefont
  {Akkermans}}\ and\ \bibinfo {author} {\bibfnamefont {G.}~\bibnamefont
  {Montambaux}},\ }\href@noop {} {\emph {\bibinfo {title} {Mesoscopic Physics
  of Electrons and Photons}}}\ (\bibinfo  {publisher} {Cambridge University
  Press},\ \bibinfo {year} {2007})\BibitemShut {NoStop}%
\bibitem [{\citenamefont {Deutsch}(2020)}]{deutsch2020harnessing}%
  \BibitemOpen
  \bibfield  {author} {\bibinfo {author} {\bibfnamefont {I.~H.}\ \bibnamefont
  {Deutsch}},\ }\href@noop {} {\bibfield  {journal} {\bibinfo  {journal} {PRX
  Quantum}\ }\textbf {\bibinfo {volume} {1}},\ \bibinfo {pages} {020101}
  (\bibinfo {year} {2020})}\BibitemShut {NoStop}%
\bibitem [{\citenamefont {Anglin}\ and\ \citenamefont
  {Ketterle}(2002)}]{anglin2002bose}%
  \BibitemOpen
  \bibfield  {author} {\bibinfo {author} {\bibfnamefont {J.~R.}\ \bibnamefont
  {Anglin}}\ and\ \bibinfo {author} {\bibfnamefont {W.}~\bibnamefont
  {Ketterle}},\ }\href@noop {} {\bibfield  {journal} {\bibinfo  {journal}
  {Nature}\ }\textbf {\bibinfo {volume} {416}},\ \bibinfo {pages} {211}
  (\bibinfo {year} {2002})}\BibitemShut {NoStop}%
\bibitem [{\citenamefont {Chen}\ \emph {et~al.}(2022)\citenamefont {Chen},
  \citenamefont {Gonz{\'a}lez~Escudero}, \citenamefont {Min{\'a}{\v{r}}},
  \citenamefont {Pasquiou}, \citenamefont {Bennetts},\ and\ \citenamefont
  {Schreck}}]{chen2022continuous}%
  \BibitemOpen
  \bibfield  {author} {\bibinfo {author} {\bibfnamefont {C.-C.}\ \bibnamefont
  {Chen}}, \bibinfo {author} {\bibfnamefont {R.}~\bibnamefont
  {Gonz{\'a}lez~Escudero}}, \bibinfo {author} {\bibfnamefont {J.}~\bibnamefont
  {Min{\'a}{\v{r}}}}, \bibinfo {author} {\bibfnamefont {B.}~\bibnamefont
  {Pasquiou}}, \bibinfo {author} {\bibfnamefont {S.}~\bibnamefont {Bennetts}},\
  and\ \bibinfo {author} {\bibfnamefont {F.}~\bibnamefont {Schreck}},\
  }\href@noop {} {\bibfield  {journal} {\bibinfo  {journal} {Nature}\ }\textbf
  {\bibinfo {volume} {606}},\ \bibinfo {pages} {683} (\bibinfo {year}
  {2022})}\BibitemShut {NoStop}%
\bibitem [{\citenamefont {Kasprzak}\ \emph {et~al.}(2006)\citenamefont
  {Kasprzak}, \citenamefont {Richard}, \citenamefont {Kundermann},
  \citenamefont {Baas}, \citenamefont {Jeambrun}, \citenamefont {Keeling},
  \citenamefont {Marchetti}, \citenamefont {Szyma{\'n}ska}, \citenamefont
  {Andr{\'e}}, \citenamefont {Staehli} \emph {et~al.}}]{kasprzak2006bose}%
  \BibitemOpen
  \bibfield  {author} {\bibinfo {author} {\bibfnamefont {J.}~\bibnamefont
  {Kasprzak}}, \bibinfo {author} {\bibfnamefont {M.}~\bibnamefont {Richard}},
  \bibinfo {author} {\bibfnamefont {S.}~\bibnamefont {Kundermann}}, \bibinfo
  {author} {\bibfnamefont {A.}~\bibnamefont {Baas}}, \bibinfo {author}
  {\bibfnamefont {P.}~\bibnamefont {Jeambrun}}, \bibinfo {author}
  {\bibfnamefont {J.~M.~J.}\ \bibnamefont {Keeling}}, \bibinfo {author}
  {\bibfnamefont {F.}~\bibnamefont {Marchetti}}, \bibinfo {author}
  {\bibfnamefont {M.}~\bibnamefont {Szyma{\'n}ska}}, \bibinfo {author}
  {\bibfnamefont {R.}~\bibnamefont {Andr{\'e}}}, \bibinfo {author}
  {\bibfnamefont {J.}~\bibnamefont {Staehli}}, \emph {et~al.},\ }\href@noop {}
  {\bibfield  {journal} {\bibinfo  {journal} {Nature}\ }\textbf {\bibinfo
  {volume} {443}},\ \bibinfo {pages} {409} (\bibinfo {year}
  {2006})}\BibitemShut {NoStop}%
\bibitem [{\citenamefont {Houck}\ \emph {et~al.}(2012)\citenamefont {Houck},
  \citenamefont {T{\"u}reci},\ and\ \citenamefont {Koch}}]{houck2012chip}%
  \BibitemOpen
  \bibfield  {author} {\bibinfo {author} {\bibfnamefont {A.~A.}\ \bibnamefont
  {Houck}}, \bibinfo {author} {\bibfnamefont {H.~E.}\ \bibnamefont
  {T{\"u}reci}},\ and\ \bibinfo {author} {\bibfnamefont {J.}~\bibnamefont
  {Koch}},\ }\href@noop {} {\bibfield  {journal} {\bibinfo  {journal} {Nature
  Physics}\ }\textbf {\bibinfo {volume} {8}},\ \bibinfo {pages} {292} (\bibinfo
  {year} {2012})}\BibitemShut {NoStop}%
\bibitem [{\citenamefont {Blatt}\ and\ \citenamefont
  {Roos}(2012)}]{blatt2012quantum}%
  \BibitemOpen
  \bibfield  {author} {\bibinfo {author} {\bibfnamefont {R.}~\bibnamefont
  {Blatt}}\ and\ \bibinfo {author} {\bibfnamefont {C.~F.}\ \bibnamefont
  {Roos}},\ }\href@noop {} {\bibfield  {journal} {\bibinfo  {journal} {Nature
  Physics}\ }\textbf {\bibinfo {volume} {8}},\ \bibinfo {pages} {277} (\bibinfo
  {year} {2012})}\BibitemShut {NoStop}%
\bibitem [{\citenamefont {Jendrzejewski}\ \emph {et~al.}(2012)\citenamefont
  {Jendrzejewski}, \citenamefont {Bernard}, \citenamefont {Mueller},
  \citenamefont {Cheinet}, \citenamefont {Josse}, \citenamefont {Piraud},
  \citenamefont {Pezz{\'e}}, \citenamefont {Sanchez-Palencia}, \citenamefont
  {Aspect},\ and\ \citenamefont {Bouyer}}]{jendrzejewski2012three}%
  \BibitemOpen
  \bibfield  {author} {\bibinfo {author} {\bibfnamefont {F.}~\bibnamefont
  {Jendrzejewski}}, \bibinfo {author} {\bibfnamefont {A.}~\bibnamefont
  {Bernard}}, \bibinfo {author} {\bibfnamefont {K.}~\bibnamefont {Mueller}},
  \bibinfo {author} {\bibfnamefont {P.}~\bibnamefont {Cheinet}}, \bibinfo
  {author} {\bibfnamefont {V.}~\bibnamefont {Josse}}, \bibinfo {author}
  {\bibfnamefont {M.}~\bibnamefont {Piraud}}, \bibinfo {author} {\bibfnamefont
  {L.}~\bibnamefont {Pezz{\'e}}}, \bibinfo {author} {\bibfnamefont
  {L.}~\bibnamefont {Sanchez-Palencia}}, \bibinfo {author} {\bibfnamefont
  {A.}~\bibnamefont {Aspect}},\ and\ \bibinfo {author} {\bibfnamefont
  {P.}~\bibnamefont {Bouyer}},\ }\href@noop {} {\bibfield  {journal} {\bibinfo
  {journal} {Nature Physics}\ }\textbf {\bibinfo {volume} {8}},\ \bibinfo
  {pages} {398} (\bibinfo {year} {2012})}\BibitemShut {NoStop}%
\bibitem [{\citenamefont {Balasubrahmaniyam}\ \emph {et~al.}(2023)\citenamefont
  {Balasubrahmaniyam}, \citenamefont {Simkhovich}, \citenamefont {Golombek},
  \citenamefont {Sandik}, \citenamefont {Ankonina},\ and\ \citenamefont
  {Schwartz}}]{balasubrahmaniyam2023enhanced}%
  \BibitemOpen
  \bibfield  {author} {\bibinfo {author} {\bibfnamefont {M.}~\bibnamefont
  {Balasubrahmaniyam}}, \bibinfo {author} {\bibfnamefont {A.}~\bibnamefont
  {Simkhovich}}, \bibinfo {author} {\bibfnamefont {A.}~\bibnamefont
  {Golombek}}, \bibinfo {author} {\bibfnamefont {G.}~\bibnamefont {Sandik}},
  \bibinfo {author} {\bibfnamefont {G.}~\bibnamefont {Ankonina}},\ and\
  \bibinfo {author} {\bibfnamefont {T.}~\bibnamefont {Schwartz}},\ }\href@noop
  {} {\bibfield  {journal} {\bibinfo  {journal} {Nature Materials}\ }\textbf
  {\bibinfo {volume} {22}},\ \bibinfo {pages} {338} (\bibinfo {year}
  {2023})}\BibitemShut {NoStop}%
\bibitem [{\citenamefont {Chen}\ \emph {et~al.}(2023)\citenamefont {Chen},
  \citenamefont {Shi}, \citenamefont {Gan}, \citenamefont {Liu}, \citenamefont
  {Li}, \citenamefont {Ghosh},\ and\ \citenamefont
  {Xiong}}]{chen2023unraveling}%
  \BibitemOpen
  \bibfield  {author} {\bibinfo {author} {\bibfnamefont {Y.}~\bibnamefont
  {Chen}}, \bibinfo {author} {\bibfnamefont {Y.}~\bibnamefont {Shi}}, \bibinfo
  {author} {\bibfnamefont {Y.}~\bibnamefont {Gan}}, \bibinfo {author}
  {\bibfnamefont {H.}~\bibnamefont {Liu}}, \bibinfo {author} {\bibfnamefont
  {T.}~\bibnamefont {Li}}, \bibinfo {author} {\bibfnamefont {S.}~\bibnamefont
  {Ghosh}},\ and\ \bibinfo {author} {\bibfnamefont {Q.}~\bibnamefont {Xiong}},\
  }\href@noop {} {\bibfield  {journal} {\bibinfo  {journal} {Nano Letters}\
  }\textbf {\bibinfo {volume} {23}},\ \bibinfo {pages} {8704} (\bibinfo {year}
  {2023})}\BibitemShut {NoStop}%
\bibitem [{\citenamefont {Carusotto}\ and\ \citenamefont
  {Ciuti}(2013)}]{carusotto2013quantum}%
  \BibitemOpen
  \bibfield  {author} {\bibinfo {author} {\bibfnamefont {I.}~\bibnamefont
  {Carusotto}}\ and\ \bibinfo {author} {\bibfnamefont {C.}~\bibnamefont
  {Ciuti}},\ }\href@noop {} {\bibfield  {journal} {\bibinfo  {journal} {Reviews
  of Modern Physics}\ }\textbf {\bibinfo {volume} {85}},\ \bibinfo {pages}
  {299} (\bibinfo {year} {2013})}\BibitemShut {NoStop}%
\bibitem [{\citenamefont {Byrnes}\ \emph {et~al.}(2014)\citenamefont {Byrnes},
  \citenamefont {Kim},\ and\ \citenamefont {Yamamoto}}]{byrnes2014exciton}%
  \BibitemOpen
  \bibfield  {author} {\bibinfo {author} {\bibfnamefont {T.}~\bibnamefont
  {Byrnes}}, \bibinfo {author} {\bibfnamefont {N.~Y.}\ \bibnamefont {Kim}},\
  and\ \bibinfo {author} {\bibfnamefont {Y.}~\bibnamefont {Yamamoto}},\
  }\href@noop {} {\bibfield  {journal} {\bibinfo  {journal} {Nature Physics}\
  }\textbf {\bibinfo {volume} {10}},\ \bibinfo {pages} {803} (\bibinfo {year}
  {2014})}\BibitemShut {NoStop}%
\bibitem [{\citenamefont {Mrejen}\ \emph {et~al.}(2019)\citenamefont {Mrejen},
  \citenamefont {Yadgarov}, \citenamefont {Levanon},\ and\ \citenamefont
  {Suchowski}}]{mrejen2019transient}%
  \BibitemOpen
  \bibfield  {author} {\bibinfo {author} {\bibfnamefont {M.}~\bibnamefont
  {Mrejen}}, \bibinfo {author} {\bibfnamefont {L.}~\bibnamefont {Yadgarov}},
  \bibinfo {author} {\bibfnamefont {A.}~\bibnamefont {Levanon}},\ and\ \bibinfo
  {author} {\bibfnamefont {H.}~\bibnamefont {Suchowski}},\ }\href@noop {}
  {\bibfield  {journal} {\bibinfo  {journal} {Science Advances}\ }\textbf
  {\bibinfo {volume} {5}},\ \bibinfo {pages} {eaat9618} (\bibinfo {year}
  {2019})}\BibitemShut {NoStop}%
\bibitem [{\citenamefont {Xu}\ \emph {et~al.}(2023)\citenamefont {Xu},
  \citenamefont {Mandal}, \citenamefont {Baxter}, \citenamefont {Cheng},
  \citenamefont {Lee}, \citenamefont {Su}, \citenamefont {Liu}, \citenamefont
  {Reichman},\ and\ \citenamefont {Delor}}]{xu2023ultrafast}%
  \BibitemOpen
  \bibfield  {author} {\bibinfo {author} {\bibfnamefont {D.}~\bibnamefont
  {Xu}}, \bibinfo {author} {\bibfnamefont {A.}~\bibnamefont {Mandal}}, \bibinfo
  {author} {\bibfnamefont {J.~M.}\ \bibnamefont {Baxter}}, \bibinfo {author}
  {\bibfnamefont {S.-W.}\ \bibnamefont {Cheng}}, \bibinfo {author}
  {\bibfnamefont {I.}~\bibnamefont {Lee}}, \bibinfo {author} {\bibfnamefont
  {H.}~\bibnamefont {Su}}, \bibinfo {author} {\bibfnamefont {S.}~\bibnamefont
  {Liu}}, \bibinfo {author} {\bibfnamefont {D.~R.}\ \bibnamefont {Reichman}},\
  and\ \bibinfo {author} {\bibfnamefont {M.}~\bibnamefont {Delor}},\
  }\href@noop {} {\bibfield  {journal} {\bibinfo  {journal} {Nature
  Communications}\ }\textbf {\bibinfo {volume} {14}},\ \bibinfo {pages} {3881}
  (\bibinfo {year} {2023})}\BibitemShut {NoStop}%
\bibitem [{\citenamefont {Ramezani}\ \emph {et~al.}(2019)\citenamefont
  {Ramezani}, \citenamefont {Halpin}, \citenamefont {Wang}, \citenamefont
  {Berghuis},\ and\ \citenamefont {Rivas}}]{ramezani2019ultrafast}%
  \BibitemOpen
  \bibfield  {author} {\bibinfo {author} {\bibfnamefont {M.}~\bibnamefont
  {Ramezani}}, \bibinfo {author} {\bibfnamefont {A.}~\bibnamefont {Halpin}},
  \bibinfo {author} {\bibfnamefont {S.}~\bibnamefont {Wang}}, \bibinfo {author}
  {\bibfnamefont {M.}~\bibnamefont {Berghuis}},\ and\ \bibinfo {author}
  {\bibfnamefont {J.~G.}\ \bibnamefont {Rivas}},\ }\href@noop {} {\bibfield
  {journal} {\bibinfo  {journal} {Nano Letters}\ }\textbf {\bibinfo {volume}
  {19}},\ \bibinfo {pages} {8590} (\bibinfo {year} {2019})}\BibitemShut
  {NoStop}%
\bibitem [{\citenamefont {Pinsker}\ \emph {et~al.}(2017)\citenamefont
  {Pinsker}, \citenamefont {Ruan},\ and\ \citenamefont
  {Alexander}}]{pinsker2017effects}%
  \BibitemOpen
  \bibfield  {author} {\bibinfo {author} {\bibfnamefont {F.}~\bibnamefont
  {Pinsker}}, \bibinfo {author} {\bibfnamefont {X.}~\bibnamefont {Ruan}},\ and\
  \bibinfo {author} {\bibfnamefont {T.}~\bibnamefont {Alexander}},\ }\href@noop
  {} {\bibfield  {journal} {\bibinfo  {journal} {Scientific reports}\ }\textbf
  {\bibinfo {volume} {7}},\ \bibinfo {pages} {1891} (\bibinfo {year}
  {2017})}\BibitemShut {NoStop}%
\bibitem [{\citenamefont {Estrecho}\ \emph {et~al.}(2019)\citenamefont
  {Estrecho}, \citenamefont {Gao}, \citenamefont {Bobrovska}, \citenamefont
  {Comber-Todd}, \citenamefont {Fraser}, \citenamefont {Steger}, \citenamefont
  {West}, \citenamefont {Pfeiffer}, \citenamefont {Levinsen}, \citenamefont
  {Parish}, \citenamefont {Liew}, \citenamefont {Matuszewski}, \citenamefont
  {Snoke}, \citenamefont {Truscott},\ and\ \citenamefont
  {Ostrovskaya}}]{EstrechoPRB2019}%
  \BibitemOpen
  \bibfield  {author} {\bibinfo {author} {\bibfnamefont {E.}~\bibnamefont
  {Estrecho}}, \bibinfo {author} {\bibfnamefont {T.}~\bibnamefont {Gao}},
  \bibinfo {author} {\bibfnamefont {N.}~\bibnamefont {Bobrovska}}, \bibinfo
  {author} {\bibfnamefont {D.}~\bibnamefont {Comber-Todd}}, \bibinfo {author}
  {\bibfnamefont {M.~D.}\ \bibnamefont {Fraser}}, \bibinfo {author}
  {\bibfnamefont {M.}~\bibnamefont {Steger}}, \bibinfo {author} {\bibfnamefont
  {K.}~\bibnamefont {West}}, \bibinfo {author} {\bibfnamefont {L.~N.}\
  \bibnamefont {Pfeiffer}}, \bibinfo {author} {\bibfnamefont {J.}~\bibnamefont
  {Levinsen}}, \bibinfo {author} {\bibfnamefont {M.~M.}\ \bibnamefont
  {Parish}}, \bibinfo {author} {\bibfnamefont {T.~C.~H.}\ \bibnamefont {Liew}},
  \bibinfo {author} {\bibfnamefont {M.}~\bibnamefont {Matuszewski}}, \bibinfo
  {author} {\bibfnamefont {D.~W.}\ \bibnamefont {Snoke}}, \bibinfo {author}
  {\bibfnamefont {A.~G.}\ \bibnamefont {Truscott}},\ and\ \bibinfo {author}
  {\bibfnamefont {E.~A.}\ \bibnamefont {Ostrovskaya}},\ }\href
  {https://doi.org/10.1103/PhysRevB.100.035306} {\bibfield  {journal} {\bibinfo
   {journal} {Physical Review B}\ }\textbf {\bibinfo {volume} {100}},\ \bibinfo
  {pages} {035306} (\bibinfo {year} {2019})}\BibitemShut {NoStop}%
\bibitem [{\citenamefont {Wu}\ \emph {et~al.}(2021)\citenamefont {Wu},
  \citenamefont {Su}, \citenamefont {Fieramosca}, \citenamefont {Ghosh},
  \citenamefont {Zhao}, \citenamefont {Liew},\ and\ \citenamefont
  {Xiong}}]{wu2021perovskite}%
  \BibitemOpen
  \bibfield  {author} {\bibinfo {author} {\bibfnamefont {J.}~\bibnamefont
  {Wu}}, \bibinfo {author} {\bibfnamefont {R.}~\bibnamefont {Su}}, \bibinfo
  {author} {\bibfnamefont {A.}~\bibnamefont {Fieramosca}}, \bibinfo {author}
  {\bibfnamefont {S.}~\bibnamefont {Ghosh}}, \bibinfo {author} {\bibfnamefont
  {J.}~\bibnamefont {Zhao}}, \bibinfo {author} {\bibfnamefont {T.~C.}\
  \bibnamefont {Liew}},\ and\ \bibinfo {author} {\bibfnamefont
  {Q.}~\bibnamefont {Xiong}},\ }\href@noop {} {\bibfield  {journal} {\bibinfo
  {journal} {Advanced Photonics}\ }\textbf {\bibinfo {volume} {3}},\ \bibinfo
  {pages} {055003} (\bibinfo {year} {2021})}\BibitemShut {NoStop}%
\bibitem [{\citenamefont {Hanai}\ \emph {et~al.}(2019)\citenamefont {Hanai},
  \citenamefont {Edelman}, \citenamefont {Ohashi},\ and\ \citenamefont
  {Littlewood}}]{Hanai2019PRL}%
  \BibitemOpen
  \bibfield  {author} {\bibinfo {author} {\bibfnamefont {R.}~\bibnamefont
  {Hanai}}, \bibinfo {author} {\bibfnamefont {A.}~\bibnamefont {Edelman}},
  \bibinfo {author} {\bibfnamefont {Y.}~\bibnamefont {Ohashi}},\ and\ \bibinfo
  {author} {\bibfnamefont {P.~B.}\ \bibnamefont {Littlewood}},\ }\href
  {https://doi.org/10.1103/PhysRevLett.122.185301} {\bibfield  {journal}
  {\bibinfo  {journal} {Physical Review Letters}\ }\textbf {\bibinfo {volume}
  {122}},\ \bibinfo {pages} {185301} (\bibinfo {year} {2019})}\BibitemShut
  {NoStop}%
\bibitem [{\citenamefont {Ghosh}\ \emph {et~al.}(2022)\citenamefont {Ghosh},
  \citenamefont {Su}, \citenamefont {Zhao}, \citenamefont {Fieramosca},
  \citenamefont {Wu}, \citenamefont {Li}, \citenamefont {Zhang}, \citenamefont
  {Li}, \citenamefont {Chen}, \citenamefont {Liew} \emph {et~al.}}]{ghoshPI22}%
  \BibitemOpen
  \bibfield  {author} {\bibinfo {author} {\bibfnamefont {S.}~\bibnamefont
  {Ghosh}}, \bibinfo {author} {\bibfnamefont {R.}~\bibnamefont {Su}}, \bibinfo
  {author} {\bibfnamefont {J.}~\bibnamefont {Zhao}}, \bibinfo {author}
  {\bibfnamefont {A.}~\bibnamefont {Fieramosca}}, \bibinfo {author}
  {\bibfnamefont {J.}~\bibnamefont {Wu}}, \bibinfo {author} {\bibfnamefont
  {T.}~\bibnamefont {Li}}, \bibinfo {author} {\bibfnamefont {Q.}~\bibnamefont
  {Zhang}}, \bibinfo {author} {\bibfnamefont {F.}~\bibnamefont {Li}}, \bibinfo
  {author} {\bibfnamefont {Z.}~\bibnamefont {Chen}}, \bibinfo {author}
  {\bibfnamefont {T.}~\bibnamefont {Liew}}, \emph {et~al.},\ }\href@noop {}
  {\bibfield  {journal} {\bibinfo  {journal} {Photonics Insights}\ }\textbf
  {\bibinfo {volume} {1}},\ \bibinfo {pages} {R04} (\bibinfo {year}
  {2022})}\BibitemShut {NoStop}%
\bibitem [{\citenamefont {Wertz}\ \emph {et~al.}(2012)\citenamefont {Wertz},
  \citenamefont {Amo}, \citenamefont {Solnyshkov}, \citenamefont {Ferrier},
  \citenamefont {Liew}, \citenamefont {Sanvitto}, \citenamefont {Senellart},
  \citenamefont {Sagnes}, \citenamefont {Lema{\^\i}tre}, \citenamefont
  {Kavokin} \emph {et~al.}}]{wertz2012propagation}%
  \BibitemOpen
  \bibfield  {author} {\bibinfo {author} {\bibfnamefont {E.}~\bibnamefont
  {Wertz}}, \bibinfo {author} {\bibfnamefont {A.}~\bibnamefont {Amo}}, \bibinfo
  {author} {\bibfnamefont {D.}~\bibnamefont {Solnyshkov}}, \bibinfo {author}
  {\bibfnamefont {L.}~\bibnamefont {Ferrier}}, \bibinfo {author} {\bibfnamefont
  {T.~C.~H.}\ \bibnamefont {Liew}}, \bibinfo {author} {\bibfnamefont
  {D.}~\bibnamefont {Sanvitto}}, \bibinfo {author} {\bibfnamefont
  {P.}~\bibnamefont {Senellart}}, \bibinfo {author} {\bibfnamefont
  {I.}~\bibnamefont {Sagnes}}, \bibinfo {author} {\bibfnamefont
  {A.}~\bibnamefont {Lema{\^\i}tre}}, \bibinfo {author} {\bibfnamefont
  {A.}~\bibnamefont {Kavokin}}, \emph {et~al.},\ }\href@noop {} {\bibfield
  {journal} {\bibinfo  {journal} {Physical Review Letters}\ }\textbf {\bibinfo
  {volume} {109}},\ \bibinfo {pages} {216404} (\bibinfo {year}
  {2012})}\BibitemShut {NoStop}%
\bibitem [{\citenamefont {Hu}\ \emph {et~al.}(2017)\citenamefont {Hu},
  \citenamefont {Luan}, \citenamefont {Scott}, \citenamefont {Yan},
  \citenamefont {Mandrus}, \citenamefont {Xu},\ and\ \citenamefont
  {Fei}}]{hu2017imaging}%
  \BibitemOpen
  \bibfield  {author} {\bibinfo {author} {\bibfnamefont {F.}~\bibnamefont
  {Hu}}, \bibinfo {author} {\bibfnamefont {Y.}~\bibnamefont {Luan}}, \bibinfo
  {author} {\bibfnamefont {M.}~\bibnamefont {Scott}}, \bibinfo {author}
  {\bibfnamefont {J.}~\bibnamefont {Yan}}, \bibinfo {author} {\bibfnamefont
  {D.}~\bibnamefont {Mandrus}}, \bibinfo {author} {\bibfnamefont
  {X.}~\bibnamefont {Xu}},\ and\ \bibinfo {author} {\bibfnamefont
  {Z.}~\bibnamefont {Fei}},\ }\href@noop {} {\bibfield  {journal} {\bibinfo
  {journal} {Nature Photonics}\ }\textbf {\bibinfo {volume} {11}},\ \bibinfo
  {pages} {356} (\bibinfo {year} {2017})}\BibitemShut {NoStop}%
\bibitem [{\citenamefont {Su}\ \emph {et~al.}(2017)\citenamefont {Su},
  \citenamefont {Diederichs}, \citenamefont {Wang}, \citenamefont {Liew},
  \citenamefont {Zhao}, \citenamefont {Liu}, \citenamefont {Xu}, \citenamefont
  {Chen},\ and\ \citenamefont {Xiong}}]{Su2017}%
  \BibitemOpen
  \bibfield  {author} {\bibinfo {author} {\bibfnamefont {R.}~\bibnamefont
  {Su}}, \bibinfo {author} {\bibfnamefont {C.}~\bibnamefont {Diederichs}},
  \bibinfo {author} {\bibfnamefont {J.}~\bibnamefont {Wang}}, \bibinfo {author}
  {\bibfnamefont {T.~C.~H.}\ \bibnamefont {Liew}}, \bibinfo {author}
  {\bibfnamefont {J.}~\bibnamefont {Zhao}}, \bibinfo {author} {\bibfnamefont
  {S.}~\bibnamefont {Liu}}, \bibinfo {author} {\bibfnamefont {W.}~\bibnamefont
  {Xu}}, \bibinfo {author} {\bibfnamefont {Z.}~\bibnamefont {Chen}},\ and\
  \bibinfo {author} {\bibfnamefont {Q.}~\bibnamefont {Xiong}},\ }\href
  {https://doi.org/10.1021/acs.nanolett.7b01956} {\bibfield  {journal}
  {\bibinfo  {journal} {Nano Letters}\ }\textbf {\bibinfo {volume} {17}},\
  \bibinfo {pages} {3982} (\bibinfo {year} {2017})}\BibitemShut {NoStop}%
\bibitem [{\citenamefont {Su}\ \emph {et~al.}(2020)\citenamefont {Su},
  \citenamefont {Ghosh}, \citenamefont {Wang}, \citenamefont {Liu},
  \citenamefont {Diederichs}, \citenamefont {Liew},\ and\ \citenamefont
  {Xiong}}]{Su2020}%
  \BibitemOpen
  \bibfield  {author} {\bibinfo {author} {\bibfnamefont {R.}~\bibnamefont
  {Su}}, \bibinfo {author} {\bibfnamefont {S.}~\bibnamefont {Ghosh}}, \bibinfo
  {author} {\bibfnamefont {J.}~\bibnamefont {Wang}}, \bibinfo {author}
  {\bibfnamefont {S.}~\bibnamefont {Liu}}, \bibinfo {author} {\bibfnamefont
  {C.}~\bibnamefont {Diederichs}}, \bibinfo {author} {\bibfnamefont {T.~C.~H.}\
  \bibnamefont {Liew}},\ and\ \bibinfo {author} {\bibfnamefont
  {Q.}~\bibnamefont {Xiong}},\ }\href
  {https://doi.org/10.1038/s41567-019-0764-5} {\bibfield  {journal} {\bibinfo
  {journal} {Nature Physics}\ }\textbf {\bibinfo {volume} {16}},\ \bibinfo
  {pages} {301} (\bibinfo {year} {2020})}\BibitemShut {NoStop}%
\bibitem [{\citenamefont {Su}\ \emph {et~al.}(2018)\citenamefont {Su},
  \citenamefont {Wang}, \citenamefont {Zhao}, \citenamefont {Xing},
  \citenamefont {Zhao}, \citenamefont {Diederichs}, \citenamefont {Liew},\ and\
  \citenamefont {Xiong}}]{su2018room}%
  \BibitemOpen
  \bibfield  {author} {\bibinfo {author} {\bibfnamefont {R.}~\bibnamefont
  {Su}}, \bibinfo {author} {\bibfnamefont {J.}~\bibnamefont {Wang}}, \bibinfo
  {author} {\bibfnamefont {J.}~\bibnamefont {Zhao}}, \bibinfo {author}
  {\bibfnamefont {J.}~\bibnamefont {Xing}}, \bibinfo {author} {\bibfnamefont
  {W.}~\bibnamefont {Zhao}}, \bibinfo {author} {\bibfnamefont {C.}~\bibnamefont
  {Diederichs}}, \bibinfo {author} {\bibfnamefont {T.~C.}\ \bibnamefont
  {Liew}},\ and\ \bibinfo {author} {\bibfnamefont {Q.}~\bibnamefont {Xiong}},\
  }\href@noop {} {\bibfield  {journal} {\bibinfo  {journal} {Science Advances}\
  }\textbf {\bibinfo {volume} {4}},\ \bibinfo {pages} {eaau0244} (\bibinfo
  {year} {2018})}\BibitemShut {NoStop}%
\bibitem [{\citenamefont {Opala}\ \emph {et~al.}(2019)\citenamefont {Opala},
  \citenamefont {Ghosh}, \citenamefont {Liew},\ and\ \citenamefont
  {Matuszewski}}]{opala2019neuromorphic}%
  \BibitemOpen
  \bibfield  {author} {\bibinfo {author} {\bibfnamefont {A.}~\bibnamefont
  {Opala}}, \bibinfo {author} {\bibfnamefont {S.}~\bibnamefont {Ghosh}},
  \bibinfo {author} {\bibfnamefont {T.~C.}\ \bibnamefont {Liew}},\ and\
  \bibinfo {author} {\bibfnamefont {M.}~\bibnamefont {Matuszewski}},\
  }\href@noop {} {\bibfield  {journal} {\bibinfo  {journal} {Physical Review
  Applied}\ }\textbf {\bibinfo {volume} {11}},\ \bibinfo {pages} {064029}
  (\bibinfo {year} {2019})}\BibitemShut {NoStop}%
\bibitem [{\citenamefont {Banerjee}\ \emph {et~al.}(2020)\citenamefont
  {Banerjee}, \citenamefont {Mandal},\ and\ \citenamefont
  {Liew}}]{banerjee2020coupling}%
  \BibitemOpen
  \bibfield  {author} {\bibinfo {author} {\bibfnamefont {R.}~\bibnamefont
  {Banerjee}}, \bibinfo {author} {\bibfnamefont {S.}~\bibnamefont {Mandal}},\
  and\ \bibinfo {author} {\bibfnamefont {T.}~\bibnamefont {Liew}},\ }\href@noop
  {} {\bibfield  {journal} {\bibinfo  {journal} {Physical Review Letters}\
  }\textbf {\bibinfo {volume} {124}},\ \bibinfo {pages} {063901} (\bibinfo
  {year} {2020})}\BibitemShut {NoStop}%
\bibitem [{\citenamefont {Nigro}\ \emph {et~al.}(2022)\citenamefont {Nigro},
  \citenamefont {D’Ambrosio}, \citenamefont {Sanvitto},\ and\ \citenamefont
  {Gerace}}]{nigro2022integrated}%
  \BibitemOpen
  \bibfield  {author} {\bibinfo {author} {\bibfnamefont {D.}~\bibnamefont
  {Nigro}}, \bibinfo {author} {\bibfnamefont {V.}~\bibnamefont {D’Ambrosio}},
  \bibinfo {author} {\bibfnamefont {D.}~\bibnamefont {Sanvitto}},\ and\
  \bibinfo {author} {\bibfnamefont {D.}~\bibnamefont {Gerace}},\ }\href@noop {}
  {\bibfield  {journal} {\bibinfo  {journal} {Communications Physics}\ }\textbf
  {\bibinfo {volume} {5}},\ \bibinfo {pages} {34} (\bibinfo {year}
  {2022})}\BibitemShut {NoStop}%
\bibitem [{\citenamefont {Xu}\ \emph {et~al.}(2021)\citenamefont {Xu},
  \citenamefont {Krisnanda}, \citenamefont {Verstraelen}, \citenamefont
  {Liew},\ and\ \citenamefont {Ghosh}}]{xu2021superpolynomial}%
  \BibitemOpen
  \bibfield  {author} {\bibinfo {author} {\bibfnamefont {H.}~\bibnamefont
  {Xu}}, \bibinfo {author} {\bibfnamefont {T.}~\bibnamefont {Krisnanda}},
  \bibinfo {author} {\bibfnamefont {W.}~\bibnamefont {Verstraelen}}, \bibinfo
  {author} {\bibfnamefont {T.~C.}\ \bibnamefont {Liew}},\ and\ \bibinfo
  {author} {\bibfnamefont {S.}~\bibnamefont {Ghosh}},\ }\href@noop {}
  {\bibfield  {journal} {\bibinfo  {journal} {Physical Review B}\ }\textbf
  {\bibinfo {volume} {103}},\ \bibinfo {pages} {195302} (\bibinfo {year}
  {2021})}\BibitemShut {NoStop}%
\bibitem [{\citenamefont {Lee}\ and\ \citenamefont
  {Ramakrishnan}(1985)}]{lee1985disordered}%
  \BibitemOpen
  \bibfield  {author} {\bibinfo {author} {\bibfnamefont {P.~A.}\ \bibnamefont
  {Lee}}\ and\ \bibinfo {author} {\bibfnamefont {T.}~\bibnamefont
  {Ramakrishnan}},\ }\href@noop {} {\bibfield  {journal} {\bibinfo  {journal}
  {Reviews of Modern Physics}\ }\textbf {\bibinfo {volume} {57}},\ \bibinfo
  {pages} {287} (\bibinfo {year} {1985})}\BibitemShut {NoStop}%
\bibitem [{\citenamefont {Wouters}\ and\ \citenamefont
  {Carusotto}(2007)}]{wouters2007excitations}%
  \BibitemOpen
  \bibfield  {author} {\bibinfo {author} {\bibfnamefont {M.}~\bibnamefont
  {Wouters}}\ and\ \bibinfo {author} {\bibfnamefont {I.}~\bibnamefont
  {Carusotto}},\ }\href@noop {} {\bibfield  {journal} {\bibinfo  {journal}
  {Physical Review Letters}\ }\textbf {\bibinfo {volume} {99}},\ \bibinfo
  {pages} {140402} (\bibinfo {year} {2007})}\BibitemShut {NoStop}%
\bibitem [{\citenamefont {Ghosh}\ \emph {et~al.}(2014)\citenamefont {Ghosh},
  \citenamefont {Cherroret}, \citenamefont {Gr{\'e}maud}, \citenamefont
  {Miniatura},\ and\ \citenamefont {Delande}}]{ghosh2014coherent}%
  \BibitemOpen
  \bibfield  {author} {\bibinfo {author} {\bibfnamefont {S.}~\bibnamefont
  {Ghosh}}, \bibinfo {author} {\bibfnamefont {N.}~\bibnamefont {Cherroret}},
  \bibinfo {author} {\bibfnamefont {B.}~\bibnamefont {Gr{\'e}maud}}, \bibinfo
  {author} {\bibfnamefont {C.}~\bibnamefont {Miniatura}},\ and\ \bibinfo
  {author} {\bibfnamefont {D.}~\bibnamefont {Delande}},\ }\href@noop {}
  {\bibfield  {journal} {\bibinfo  {journal} {Physical Review A}\ }\textbf
  {\bibinfo {volume} {90}},\ \bibinfo {pages} {063602} (\bibinfo {year}
  {2014})}\BibitemShut {NoStop}%
\bibitem [{\citenamefont {Anderson}(1958)}]{anderson1958absence}%
  \BibitemOpen
  \bibfield  {author} {\bibinfo {author} {\bibfnamefont {P.~W.}\ \bibnamefont
  {Anderson}},\ }\href@noop {} {\bibfield  {journal} {\bibinfo  {journal}
  {Physical Review}\ }\textbf {\bibinfo {volume} {109}},\ \bibinfo {pages}
  {1492} (\bibinfo {year} {1958})}\BibitemShut {NoStop}%
\bibitem [{\citenamefont {Pikovsky}\ and\ \citenamefont
  {Shepelyansky}(2008)}]{Shepelyansky08}%
  \BibitemOpen
  \bibfield  {author} {\bibinfo {author} {\bibfnamefont {A.~S.}\ \bibnamefont
  {Pikovsky}}\ and\ \bibinfo {author} {\bibfnamefont {D.~L.}\ \bibnamefont
  {Shepelyansky}},\ }\href {https://doi.org/10.1103/PhysRevLett.100.094101}
  {\bibfield  {journal} {\bibinfo  {journal} {Physical Review Letters}\
  }\textbf {\bibinfo {volume} {100}},\ \bibinfo {pages} {094101} (\bibinfo
  {year} {2008})}\BibitemShut {NoStop}%
\bibitem [{\citenamefont {Abrahams}\ \emph {et~al.}(1979)\citenamefont
  {Abrahams}, \citenamefont {Anderson}, \citenamefont {Licciardello},\ and\
  \citenamefont {Ramakrishnan}}]{GangOf4}%
  \BibitemOpen
  \bibfield  {author} {\bibinfo {author} {\bibfnamefont {E.}~\bibnamefont
  {Abrahams}}, \bibinfo {author} {\bibfnamefont {P.~W.}\ \bibnamefont
  {Anderson}}, \bibinfo {author} {\bibfnamefont {D.~C.}\ \bibnamefont
  {Licciardello}},\ and\ \bibinfo {author} {\bibfnamefont {T.~V.}\ \bibnamefont
  {Ramakrishnan}},\ }\href {https://doi.org/10.1103/PhysRevLett.42.673}
  {\bibfield  {journal} {\bibinfo  {journal} {Physical Review Letters}\
  }\textbf {\bibinfo {volume} {42}},\ \bibinfo {pages} {673} (\bibinfo {year}
  {1979})}\BibitemShut {NoStop}%
\bibitem [{\citenamefont {Thouless}(1974)}]{thouless1974electrons}%
  \BibitemOpen
  \bibfield  {author} {\bibinfo {author} {\bibfnamefont {D.~J.}\ \bibnamefont
  {Thouless}},\ }\href@noop {} {\bibfield  {journal} {\bibinfo  {journal}
  {Physics Reports}\ }\textbf {\bibinfo {volume} {13}},\ \bibinfo {pages} {93}
  (\bibinfo {year} {1974})}\BibitemShut {NoStop}%
\bibitem [{\citenamefont {Cherroret}\ \emph {et~al.}(2014)\citenamefont
  {Cherroret}, \citenamefont {Vermersch}, \citenamefont {Garreau},\ and\
  \citenamefont {Delande}}]{cherroret2014nonlinear}%
  \BibitemOpen
  \bibfield  {author} {\bibinfo {author} {\bibfnamefont {N.}~\bibnamefont
  {Cherroret}}, \bibinfo {author} {\bibfnamefont {B.}~\bibnamefont
  {Vermersch}}, \bibinfo {author} {\bibfnamefont {J.~C.}\ \bibnamefont
  {Garreau}},\ and\ \bibinfo {author} {\bibfnamefont {D.}~\bibnamefont
  {Delande}},\ }\href@noop {} {\bibfield  {journal} {\bibinfo  {journal}
  {Physical Review Letters}\ }\textbf {\bibinfo {volume} {112}},\ \bibinfo
  {pages} {170603} (\bibinfo {year} {2014})}\BibitemShut {NoStop}%
\bibitem [{\citenamefont {Vollhardt}\ and\ \citenamefont
  {W{\"o}lfle}(1982)}]{vollhardt1982scaling}%
  \BibitemOpen
  \bibfield  {author} {\bibinfo {author} {\bibfnamefont {D.}~\bibnamefont
  {Vollhardt}}\ and\ \bibinfo {author} {\bibfnamefont {P.}~\bibnamefont
  {W{\"o}lfle}},\ }\href@noop {} {\bibfield  {journal} {\bibinfo  {journal}
  {Physical Review Letters}\ }\textbf {\bibinfo {volume} {48}},\ \bibinfo
  {pages} {699} (\bibinfo {year} {1982})}\BibitemShut {NoStop}%
\bibitem [{\citenamefont {Cherroret}(2016)}]{cherroret16}%
  \BibitemOpen
  \bibfield  {author} {\bibinfo {author} {\bibfnamefont {N.}~\bibnamefont
  {Cherroret}},\ }\href@noop {} {\bibfield  {journal} {\bibinfo  {journal}
  {Journal of Physics: Condensed Matter}\ }\textbf {\bibinfo {volume} {29}},\
  \bibinfo {pages} {024002} (\bibinfo {year} {2016})}\BibitemShut {NoStop}%
\bibitem [{\citenamefont {Prat}\ \emph {et~al.}(2019)\citenamefont {Prat},
  \citenamefont {Delande},\ and\ \citenamefont {Cherroret}}]{prat2019quantum}%
  \BibitemOpen
  \bibfield  {author} {\bibinfo {author} {\bibfnamefont {T.}~\bibnamefont
  {Prat}}, \bibinfo {author} {\bibfnamefont {D.}~\bibnamefont {Delande}},\ and\
  \bibinfo {author} {\bibfnamefont {N.}~\bibnamefont {Cherroret}},\ }\href@noop
  {} {\bibfield  {journal} {\bibinfo  {journal} {Physical Review A}\ }\textbf
  {\bibinfo {volume} {99}},\ \bibinfo {pages} {023629} (\bibinfo {year}
  {2019})}\BibitemShut {NoStop}%
\bibitem [{\citenamefont {Sajjad}\ \emph {et~al.}(2022)\citenamefont {Sajjad},
  \citenamefont {Tanlimco}, \citenamefont {Mas}, \citenamefont {Cao},
  \citenamefont {Nolasco-Martinez}, \citenamefont {Simmons}, \citenamefont
  {Santos}, \citenamefont {Vignolo}, \citenamefont {Macr{\`\i}},\ and\
  \citenamefont {Weld}}]{sajjad2022observation}%
  \BibitemOpen
  \bibfield  {author} {\bibinfo {author} {\bibfnamefont {R.}~\bibnamefont
  {Sajjad}}, \bibinfo {author} {\bibfnamefont {J.~L.}\ \bibnamefont
  {Tanlimco}}, \bibinfo {author} {\bibfnamefont {H.}~\bibnamefont {Mas}},
  \bibinfo {author} {\bibfnamefont {A.}~\bibnamefont {Cao}}, \bibinfo {author}
  {\bibfnamefont {E.}~\bibnamefont {Nolasco-Martinez}}, \bibinfo {author}
  {\bibfnamefont {E.~Q.}\ \bibnamefont {Simmons}}, \bibinfo {author}
  {\bibfnamefont {F.~L.}\ \bibnamefont {Santos}}, \bibinfo {author}
  {\bibfnamefont {P.}~\bibnamefont {Vignolo}}, \bibinfo {author} {\bibfnamefont
  {T.}~\bibnamefont {Macr{\`\i}}},\ and\ \bibinfo {author} {\bibfnamefont
  {D.~M.}\ \bibnamefont {Weld}},\ }\href@noop {} {\bibfield  {journal}
  {\bibinfo  {journal} {Physical Review X}\ }\textbf {\bibinfo {volume} {12}},\
  \bibinfo {pages} {011035} (\bibinfo {year} {2022})}\BibitemShut {NoStop}%
\bibitem [{\citenamefont {Janarek}\ \emph {et~al.}(2020)\citenamefont
  {Janarek}, \citenamefont {Delande}, \citenamefont {Cherroret},\ and\
  \citenamefont {Zakrzewski}}]{janarek2020quantum}%
  \BibitemOpen
  \bibfield  {author} {\bibinfo {author} {\bibfnamefont {J.}~\bibnamefont
  {Janarek}}, \bibinfo {author} {\bibfnamefont {D.}~\bibnamefont {Delande}},
  \bibinfo {author} {\bibfnamefont {N.}~\bibnamefont {Cherroret}},\ and\
  \bibinfo {author} {\bibfnamefont {J.}~\bibnamefont {Zakrzewski}},\
  }\href@noop {} {\bibfield  {journal} {\bibinfo  {journal} {Physical Review
  A}\ }\textbf {\bibinfo {volume} {102}},\ \bibinfo {pages} {013303} (\bibinfo
  {year} {2020})}\BibitemShut {NoStop}%
\end{thebibliography}%

\end{document}